# In-depth study of spectroscopic properties of new $Pr^{3+}$-ion doped low-phonon sesquisulfide $Lu_2S_3$ material for mid-IR laser sources

Martin Fibrich [a,c,*], Jan Šulc [a], Lubomír Havlák [b], Vítězslav Jarý [b], Robert Král [b], Vojtěch Vaněček [b], David Vyhlídal [a], Helena Jelínková [a], Martin Nikl [b]

[a] Faculty of Nuclear Sciences and Physical Engineering, Czech Technical University in Prague, Břehová 7, Prague 1 115 19, Czech Republic
[b] Institute of Physics of the Czech Academy of Science, Cukrovarnická 10, Prague 6 162 00, Czech Republic
[c] The Extreme Light Infrastructure ERIC, ELI Beamlines Facility, Za Radnicí 835, Dolní Břežany 252 41, Czech Republic

ARTICLE INFO

ABSTRACT

$Lu_2S_3$ material from the family of sesquisulfide hosts appears to be a promising low-phonon material for use as a laser gain medium allowing for a laser emission over a broad spectral range from ultraviolet to far mid-infrared wavelengths. In this paper, a praseodymium-ion doped $Lu_2S_3$ single crystal grown by micro-pulling down technique is investigated with focus on its spectroscopic properties. The Raman, excitation, and luminescence spectra are presented. By selective excitation of the main energy levels of $Pr^{3+}$-ions, 26 luminescence transitions of the Pr:$Lu_2S_3$ crystal spanning the ~0.49 – 5.5 μm wavelength range have been identified. The correct assignment of the observed luminescence spectra to the respective $Pr^{3+}$ energy-level transitions was confirmed by the calculation of the optimized squared reduced-matrix elements for the tensor operators $\mathbf{U}^{(k)}$ and $\mathbf{L} + g\mathbf{S}$.

## 1. Introduction

The search for low-phonon laser materials doped with rare-earth (RE) ions is of particular interest for the development of mid-infrared (mid-IR) laser sources. Mid-IR solid-state lasers are attractive due to their wide range of potential applications, including remote sensing, molecular spectroscopy, atmospheric sensing, optical metrology, medicine, etc. Compared to optical parametric oscillators (OPOs) which are commonly used to cover this spectral range, solid-state lasers based on the direct generation of the mid-IR radiation offer greater sturdiness and simplicity.

To achieve laser emission in the mid-IR spectral region, it is crucial to select a host material with a low phonon energy. Lower phonon energies reduce the probability of multi-phonon relaxation processes of excited states. This leads to lower non-radiative losses and higher fluorescence quantum yields [1,2]. The maximum phonon energies determine the minimum number of phonons required for multi-phonon transitions between two energy levels with a given energy gap. A general guideline suggests that the laser transition energy should exceed five times the maximum phonon energy to avoid multiphonon-assisted relaxation [3]. Following this guideline, one can deduce that a laser host with a maximum phonon energy of $E_{ph} \leq 300$ cm$^{-1}$ is necessary for achieving an acceptable quantum efficiency for the longest ever reported laser transition of ~7 μm (1500 cm$^{-1}$) [4].

Efficient pulsed laser operation at the longest reported wavelength of 7.2 μm has been demonstrated in $Pr^{3+}$-ion doped $LaCl_3$ [4]. This material has a maximum phonon energy of only 210 cm$^{-1}$. However, the main drawback of such simple tri-chloride crystals, like $LaCl_3$, $ErCl_3$, $YCl_3$, etc., is that these materials suffer poor mechanical properties and requires special handling and storage due to its strong hygroscopicity [5,6]. More practical low-phonon laser materials are RE-ion doped alkali-lead halides with general formula of $MPb_2X_5$ (M = Rb, K; X = Cl, Br), having the lowest phonon energies ($E_{ph}$ ~ 140 – 200 cm$^{-1}$) ever reported in crystalline laser matrices [1–3,7,8]. These crystals are low- or non-hygroscopic, have satisfactory mechanical properties, and are able to incorporate RE ions fairly easily, although the high concentrations are difficult to achieve [1,9]. Laser emission was successfully demonstrated with $KPb_2Cl_5$ doped by $Pr^{3+}$, $Nd^{3+}$, $Dy^{3+}$, or $Er^{3+}$ ions [10–13,14]. Suitable and well-developed low-phonon host materials for mid-IR laser generation are also chalcogenide crystals like $Cr^{2+}$ and $Fe^{2+}$

* Corresponding author at: Faculty of Nuclear Sciences and Physical Engineering, Czech Technical University in Prague, Brehová 7, Prague 1 115 19, Czech Republic.

*E-mail addresses:* martin.fibrich@fjfi.cvut.cz (M. Fibrich), jan.sulc@fjfi.cvut.cz (J. Šulc), havlak@fzu.cz (L. Havlák), jary@fzu.cz (V. Jarý), kral@fzu.cz (R. Král), vanecek@fzu.cz (V. Vaněček), david.vyhlidal@fjfi.cvut.cz (D. Vyhlídal), helena.jelinkova@fjfi.cvut.cz (H. Jelínková), nikl@fzu.cz (M. Nikl).

doped II-VI semiconductor compounds [15–18–21] with typical representatives of ZnSe or ZnS ($E_{ph} \sim 200 - 350$ cm$^{-1}$) or thiogallates $CaGa_2S_4$ or $PbGa_2S_4$, for which laser action was obtained under $Dy^{3+}$ doping [10,22–24]. The great advantage of these materials is their low or zero hygroscopicity. However, they usually have low acceptance of RE-dopants, which limits the doping concentration to a few tenths of a percent [3]. New promising moisture-resistant low-phonon ($E_{ph} \sim 220$ cm$^{-1}$) chalcogenide materials, tolerant to the concentration of dopants from the lanthanide group are the ternary alkali RE sulfides with general formula $ALnS_2$ (A = K, Rb; Ln = La, Gd, Lu), represented e.g. by $KLuS_2$ material [25–28]. However, the preparation of this material is challenging, and current methods can only synthesize crystals in the form of 40 $\mu$m thick hexagonal plates [28]. This fact restricts the practical use of this material as an active medium for mid-IR laser sources.

Another interesting family of low-phonon laser hosts is the family of binary sulfides, also known as sesquisulfides for the composition $RE_2S_3$ where *RE* is the rare-earth element and *S* is sulfur. The sesquisulfides of the trivalent RE are a large family of high bandgap semiconductors with potential applications as optical IR window materials, thermoelectric and magneto-optical materials, photocatalysts, and laser phosphors, etc. [29–32,33]. These materials have several unique properties such as a high refractive index (2 – 2.8), a broad transparency range (0.5 – 20 $\mu$m), and a wide region of homogeneity of physical properties [29, 34].

This contribution focuses on the $Lu_2S_3$ crystal, which has been successfully grown in bulk form. The material is highly stable and exhibits chemical resistivity comparable to that of ZnSe [35]. However, unlike II-VI semiconductor compounds, its structure makes it highly suitable for lanthanide doping. Thus, the RE-doped sesquisulfides have the potential to be an interesting alternative to the well-developed ZnS and ZnSe materials, with the aim of extending the range of available laser wavelengths in the mid-IR region. This article describes the spectroscopic properties of $Pr^{3+}$ ion-doped $Lu_2S_3$ crystals, successively grown by micro-pulling-down method [36]. The main focus is on excitation and luminescence spectra which are firstly (to our best knowledge) described up to 5.4 $\mu$m. Raman spectroscopy is also presented, confirming the low-phonon energy of the $Lu_2S_3$ matrix, comparable to the well-established solid-state active media in the mid-IR wavelength region, such as ZnS, ZnSe, $PbGa_2S_4$ etc.

It should be noted that $Pr^{3+}$-ions are well-established RE dopants for solid state laser hosts operating mainly in the wide spectral range in the visible region [37], but the $Pr^{3+}$ energy level structure allows for a possible laser transitions in the near- and mid-IR spectral regions. Therefore, the knowledge of the luminescence spectra up to far mid-IR is justified. The simplified energy level diagram of $Pr^{3+}$-ion along with the observed luminescence transitions of the $Pr:Lu_2S_3$ is displayed in Fig. 1.

## 2. Material and methods

### 2.1. Sample preparation

$Pr:Lu_2S_3$ crystals were grown from melt using a modified micro-pulling-down method under an inert argon atmosphere in a graphite crucible and multilayer alumina shielding. Such arrangement was necessary due to high melting temperature of $Lu_2S_3$ reaching ~1750°C. After melting the material and seeding, the $Lu_2S_3$ crystals were pulled with growth rate around 0.1 mm/min. In this way, $Pr:Lu_2S_3$ crystals with $Pr^{3+}$-ion concentration of 0.05 at. % Pr/Lu and 0.5 at. % in melt Pr/Lu were prepared. Due to the small size (area) of the samples and relatively low $Pr^{3+}$ concentrations in the melt, accurate measurement of $Pr^{3+}$ content in the sample proved to be challenging. However, X-ray fluorescence analysis of the Pr(0.5 at. %):$Lu_2S_3$ sample estimated the $Pr^{3+}$ concentration to be around 0.3 at. % Pr/Lu, closely matching the nominal melt composition. For Pr(0.05 at. %):$Lu_2S_3$, it was not possible to measure it. Therefore, hereafter we will refer to the concentration in the melt. It is worth mentioning, the distribution of $Pr^{3+}$ will be influenced by segregation in both axial and radial directions. However, given the growth conditions used for the micro-pulling-down growth of $Lu_2S_3$, segregation is expected to be significantly reduced compared to crystals grown using the Czochralski technique [38,39]. Accurate determination of the effective segregation coefficient of $Pr^{3+}$ in $Lu_2S_3$, both axially and radially, will be addressed in future studies.

It should be also noted that due to the unavailability of high-quality $Lu_2S_3$ single crystals to serve as seeds, a graphite rod was utilized instead. This approach is standard practice during the initial growth stages of novel materials. Subsequently, the $Lu_2S_3$ crystal obtained from the graphite seed can itself be employed as a seed crystal for further growth cycles. Detail description of $Pr:Lu_2S_3$ crystals growth is shown in [36].

Samples for spectroscopic characterization were prepared by cutting and polishing the grown bulk material into ~1.5 mm in diameter and ~1 mm in length pieces.

A photograph of the $Pr:Lu_2S_3$ sample (0.05 at. % of $Pr^{3+}$) is shown in Fig. 2. The sample is transparent, however, it contains black inclusions in some parts. These inclusions are most likely graphite that is released from the hot zone elements during the growth process. To avoid this phenomenon, the use of crucible made from alternative materials must be considered. However, selecting an appropriate crucible for the growth of high-temperature sulfides presents significant challenges. High-melting-point metals such as molybdenum, tungsten, or even iridium are not chemically inert toward sulfide melts at elevated temperatures. While graphite exhibits chemical resistance to sulfide melts, it tends to release particulate at high temperatures. To mitigate this issue and enhance crystal quality, we plan to utilize crucibles made from glassy carbon. However, fabricating custom-made glassy carbon crucibles of such small dimensions involves specialized manufacturing processes, making their production both time-consuming and costly.

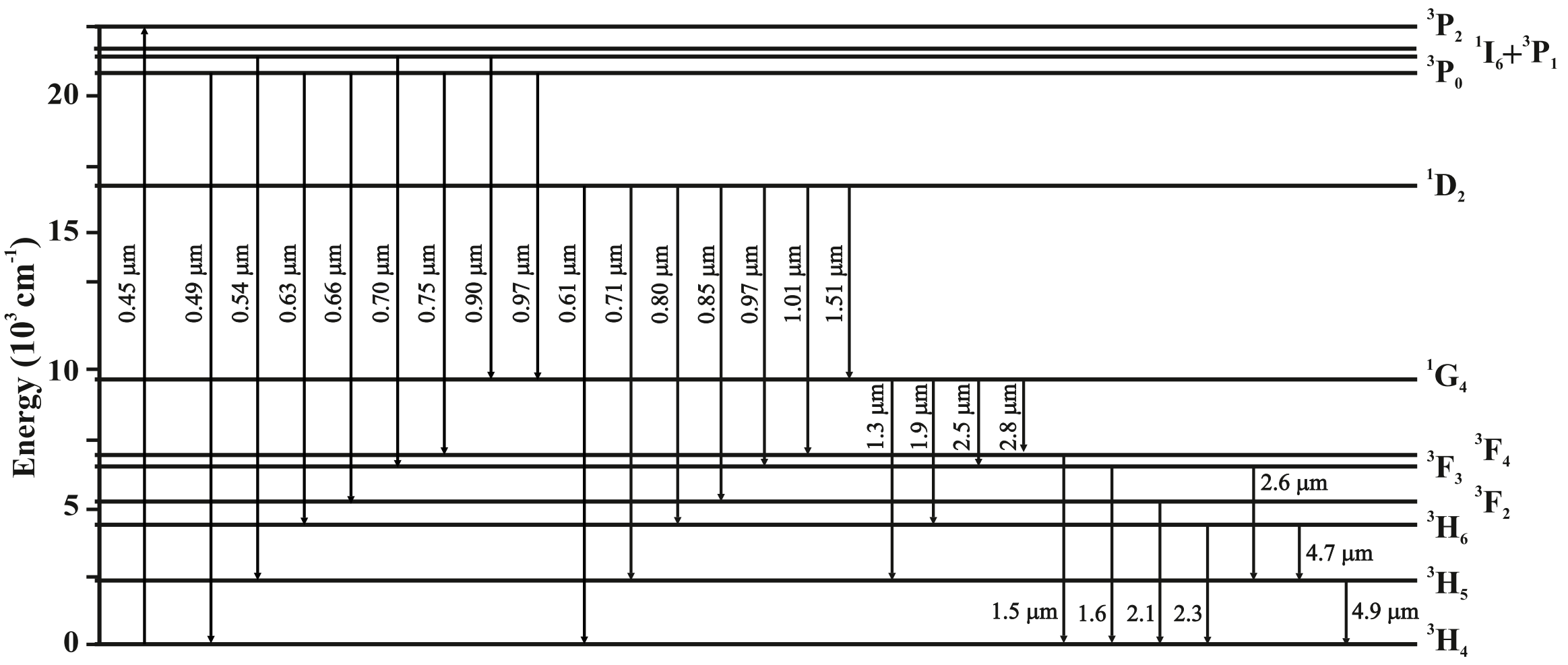

**Fig. 1.** Simplified energy level structure of $Pr^{3+}$ ions with indication of observed and identified $Pr^{3+}$ luminescence transitions in $Lu_2S_3$ laser host.

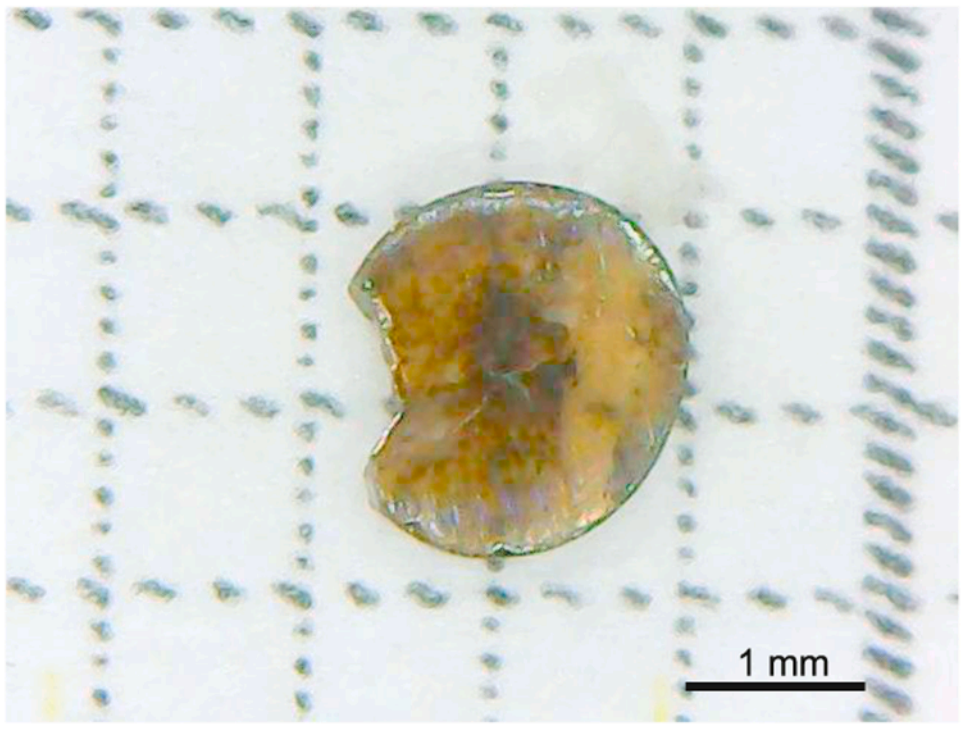

**Fig. 2.** Photograph of Pr:$Lu_2S_3$ sample (0.05 at. % of $Pr^{3+}$).

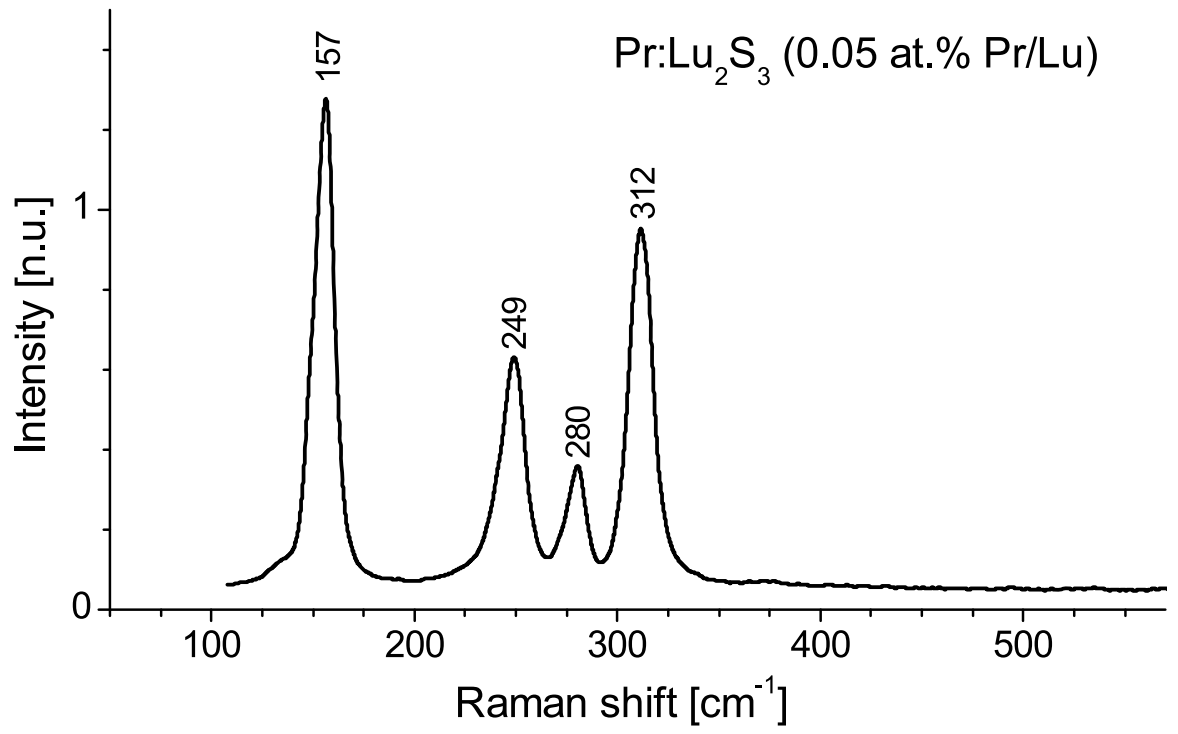

**Fig. 3.** Raman spectrum of Pr:$Lu_2S_3$ (0.05 at. % of $Pr^{3+}$) crystal.

In addition, based on our experience with the micro-pulling-down growth of oxide materials, we anticipate the formation of sulfur vacancies in the crystals when these are grown under an inert atmosphere. This issue could be mitigated by employing an argon-hydrogen sulfide gas mixture as the growth atmosphere. However, due to the toxicity of hydrogen sulfide, strict safety precautions must be implemented. Additionally, other defects, such as oxygen substitution at sulfur sites, may also occur in the resulting crystals.

### 2.2. Methods of characterization

The Raman spectrum of the Pr:$Lu_2S_3$ (0.05 at. % of $Pr^{3+}$) sample was measured using a Raman spectrometer Wasatch Photonics WP830 (Raman spectral range 100 – 1800 $cm^{-1}$, resolution 6 $cm^{-1}$).

The absorption spectra of the sample are typically measured directly using a spectrophotometer. However, direct measurement was hindered by the presence of black inclusions in our samples (see Fig. 2), which could potentially lead to distorted results. Therefore, to identify the absorption maxima of the Pr:$Lu_2S_3$, the excitation spectra measurement was performed. The investigated sample was excited by a tunable laser Ekspla NT252–1k-SH and the corresponding intensities of the fluorescence spectra were recorded as a function of the excitation wavelengths using appropriate detectors (Thorlabs Si-detector FDS10X10, PbS-detector PDA30G-EC, or PbSe-detector PDA20H-EC) in combination with suitable band-pass or long-pass filters.

The sample was then excited to its specific absorption maxima and the luminescence spectra were measured using the StelarNet BlueWave NIR-50 fiber coupled spectrometer (in the range of 490 – 1100 nm), StelarNet DwarfStar fiber coupled spectrometer (970 – 1700 nm), and stepper-motor-driven monochromator ORIEL 77250 (1900 – 5600 nm). The monochromator was equipped with a 77301 grating assembly (Newport, 150 grooves per mm) and micrometer-adjustable slits. To record the transmitted radiation intensity, the above mentioned PbS or PbSe detectors were used along with the appropriate long-pass filters to block signals from undesired diffraction orders.

## 3. Results and discussion

### 3.1. Raman spectra

The Raman spectrum of the Pr(0.05 at. %):$Lu_2S_3$ single crystal in the energy range of 100 – 550 $cm^{-1}$ is depicted in Fig. 3. The peak with a maximum at 157 $cm^{-1}$ probably corresponds to low-frequency translational modes of Lu and S in the $Lu_2S_3$ lattice (deformation vibrations). The peak at 249 $cm^{-1}$ can then be related to the bending of the Lu-S bonds. The peak at 280 $cm^{-1}$ is probably the symmetric stretching mode $A_g$, while the peak at 312 $cm^{-1}$ can be identified as the antisymmetric stretching mode $E_g$. The highest observed energy peak position can be considered as a first approximation for the effective phonon energy of the crystalline matrices [8]. Therefore, in the case of Pr:$Lu_2S_3$, the maximum value of the phonon energy can be estimated to be $E_{ph} \sim 312$ $cm^{-1}$, which is a value comparable to other low-phonon laser hosts like ZnSe ($E_{ph} \sim 250$ $cm^{-1}$), ZnS ($E_{ph} \sim 350$ $cm^{-1}$), or $PbGa_2S_4$ ($E_{ph} \sim 350$ $cm^{-1}$).

### 3.2. Excitation spectra

Due to the nature of the samples, it was not possible to directly measure absorption spectra of the crystal to set the optimal wavelengths for particular energy-level excitations, as was mentioned in 2.2. For that reason, to determine the position of the absorption maxima for individual $Pr^{3+}$-ion transitions, the excitation spectra were measured. The excitation spectra of the Pr:$Lu_2S_3$ crystal (0.05 at. % of $Pr^{3+}$) are presented in Fig. 4 for the spectral range of 440 – 515 nm (Fig. 4a) and 590 – 635 nm (Fig. 4b). Sharp absorption lines typical for the $Pr^{3+}$-ion can be identified in the distinctive laser transitions of $^3H_4 \rightarrow {}^3P_{0,1,2}$ and $^3H_4 \rightarrow {}^1D_2$, respectively. The main absorption maxima correspond to wavelengths of 456, 498, and 595 nm.

In the case of excitation spectra measurement in the infrared spectral range, a sample with a higher doping of active ions (0.5 at. % of $Pr^{3+}$) was used. This was done to increase the absorption of the excitation radiation in the crystal, resulting in a stronger signal compared to the lower doped sample. The recorded excitation spectra related to infrared absorption are shown in Fig. 5. The obtained spectra in the ranges of 960–1110 nm, 1380 – 1750 nm, and 1900 – 2230 nm correspond to $^3H_4 \rightarrow {}^1G_4$ (Fig. 5a), $^3H_4 \rightarrow {}^3F_{3,4}$ (Fig. 5b), and $^3H_4 \rightarrow {}^3F_2$ (Fig. 5c) transitions, respectively. The main absorption maxima, suitable for the further fluorescence excitation, correspond to wavelengths of 1022, 1556, 1638, and 2057 nm. The excitation spectrum for the $^3H_4 \rightarrow {}^3H_6$ transition was not measured due to the low output energy of the excitation laser for the wavelengths above 2.3 $\mu$m.

### 3.3. Pr:$Lu_2S_3$ wavefunctions optimization

Optimization of the 4 f wavefunctions of rare-earth doped materials is essential to determine the relevant values of the reduced matrix elements of the $\mathbf{U}^{(k)}(k = 2, 4, 6)$. This tensor characterizes the electric-dipole induced 4f-4f transition intensities. Tabulated values of $|\langle \mathbf{U}^{(k)} \rangle|^2$, often based on aqueous $RE^{3+}$ solutions, can differ from the optimized values for a specific matrix by up to tens of percent [40]. This discrepancy is subsequently transferred directly to the determination of the Judd-Ofelt intensity parameters. Therefore, without finding the optimized 4 f wavefunction or $|\langle \mathbf{U}^{(k)} \rangle|^2$, it is not possible to accurately calculate the intensities of absorption and emission transitions.

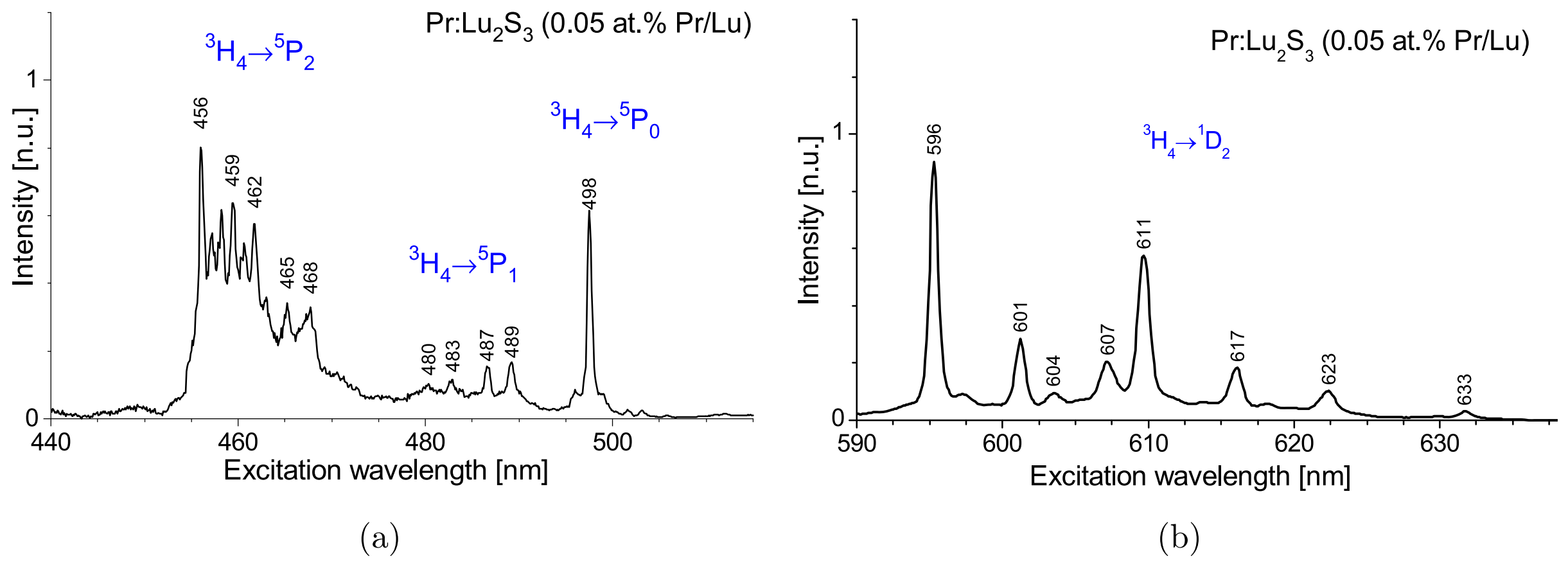

**Fig. 4.** Excitation spectra of 0.05% Pr:$Lu_2S_3$ for the spectral range of 440 – 530 nm (a), and 590 – 635 nm (b).

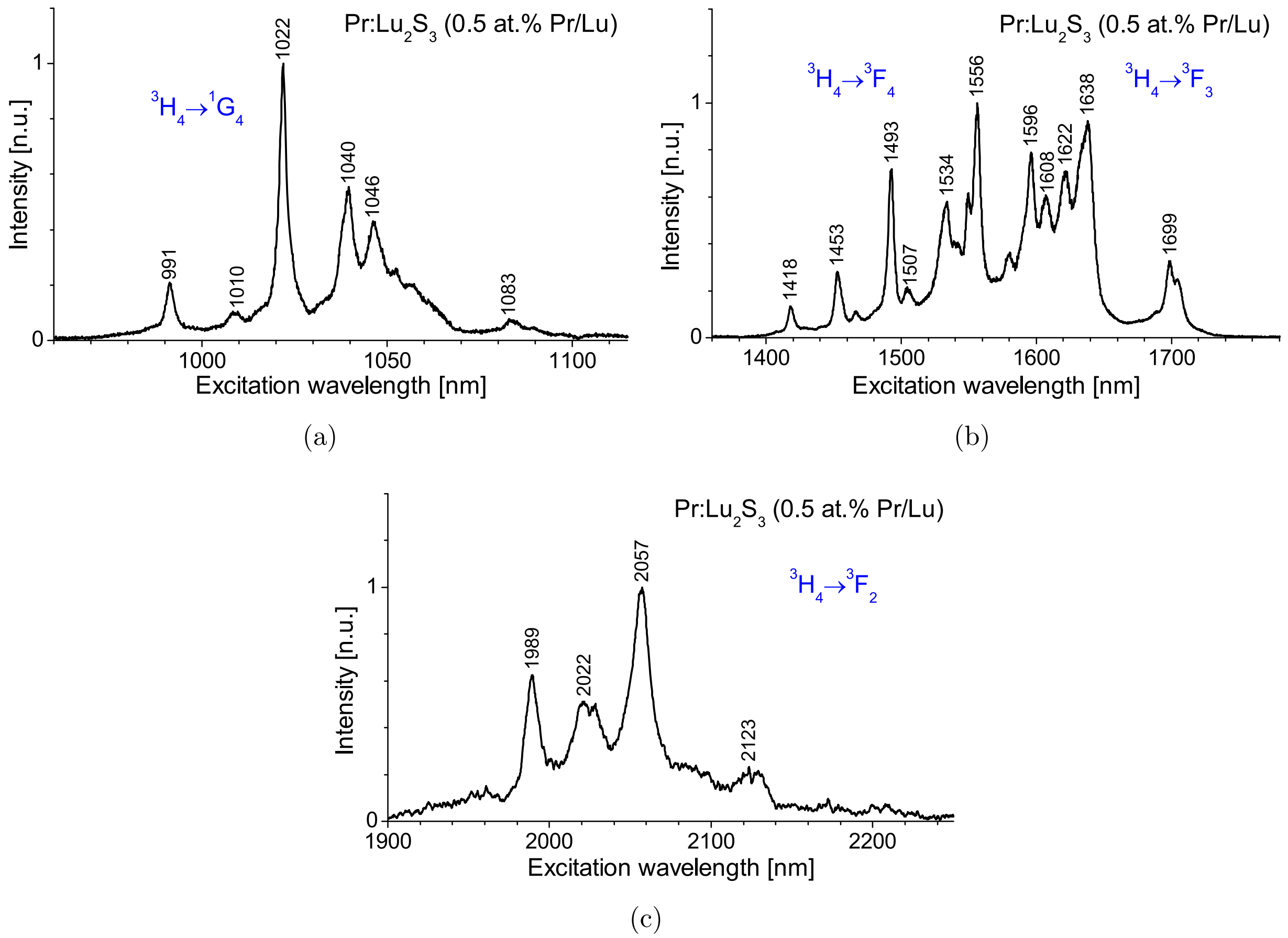

**Fig. 5.** Excitation spectra of 0.5% Pr:$Lu_2S_3$ for the spectral range of 960 – 1115 nm (a), 1380 – 1780 nm (b), and 1900 – 2280 nm (c).

To optimize the 4f wavefunctions of the investigated Pr:$Lu_2S_3$, the energies of the $^{2S+1}L_J$ multiplets must be known. These energies can be firstly determined as the barycenters of the respective transitions using the measured excitation spectra. The barycenter energy is then set as the weighted average energy for the specific transition, meaning that the integrated portions of the spectrum below and above this energy are equal. The determined barycenter energy values, along with the integration intervals, are listed in the Table 1. It should be noted that the determined barycenter values for the $^3H_4 \rightarrow {}^3F_{3,4}$ transitions may be subject to error due to the potential overlap of the bands and the uncertainty associated with the margin between the bands. It should be noted that although low-temperature excitation spectra measurements could help refine the determination of some $^{2S+1}L_J$ sublevel positions, a complete crystal-field analysis would be required to accurately determine the barycenter energies, which is laborious and not trivial task. Additionally, the improvement in the accuracy of 4f wave function optimization achieved through a detailed crystal-field analysis, compared to calculations based on room-temperature spectra measurement, is relatively small. For example, as demonstrated by Hehlen et al. [40] for Er:$LaCl_3$, using precise $^{2S+1}L_J$ multiplet energies (rather than values derived from room-temperature absorption spectra measurement) reduces the relative error in the barycenter determination, with an average energy devition of about 40 $cm^{-1}$. Even for the lowest energy levels, this corresponds to a relative error of less than 0.5 %. Therefore, we believe that our approach, based on room-temperature data, still provides a significant improvement over calculations relying on absorption spectra measurements of aqueous $RE^{3+}$ solutions.

**Table 1**
Input data for $Pr:Lu_2S_3$ wavefunctions optimization: integration intervals for barycenter determination; experimental barycenters obtained from the measured excitation spectra; barycenters calculated from optimized Hamiltonian matrix; calculated complete set of intermediate coupling wavefunctions.

| $Pr^{3+}$ level | Integration range [$cm^{-1}$] | Barycenter exp. [$cm^{-1}$] | Barycenter calc. [$cm^{-1}$] | Intermediate coupling wavefunction |
|---|---|---|---|---|
| $^3H_4$ | | | 0 | $-0.985\lvert ^3H_4\rangle - 0.168\lvert ^1G_4\rangle$ |
| $^3H_5$ | – | – | 2077 | $+1.000\lvert ^3H_5\rangle$ |
| $^3H_6$ | – | – | 4242 | $-0.998\lvert ^3H_6\rangle + 0.057\lvert ^1I_6\rangle$ |
| $^3F_2$ | 4400 – 5300 | 4888 | 4687 | $-0.988\lvert ^3F_2\rangle - 0.151\lvert ^1D_2\rangle$ |
| $^3F_3$ | 5800 – 6365 | 6162 | 6073 | $+1.000\lvert ^3F_3\rangle$ |
| $^3F_4$ | 6365 – 7100 | 6494 | 6529 | $+0.804\lvert ^3F_4\rangle - 0.581\lvert ^1G_4\rangle + 0.124\lvert ^3H_4\rangle$ |
| $^1G_4$ | 8500 – 11,000 | 9639 | 9450 | $+0.796\lvert ^1G_4\rangle + 0.593\lvert ^3F_4\rangle - 0.117\lvert ^3H_4\rangle$ |
| $^1D_2$ | 15,000 – 17,200 | 16,479 | 16,450 | $-0.944\lvert ^1D_2\rangle + 0.295\lvert ^3P_2\rangle + 0.149\lvert ^3F_2\rangle$ |
| $^3P_0$ | 19,800 – 20,280 | 20,099 | 20,112 | $-0.996\lvert ^3P_0\rangle - 0.094\lvert ^1S_0\rangle$ |
| $^1I_6$ | – | – | 20,124 | $+0.998\lvert ^1I_6\rangle + 0.057\lvert ^3H_6\rangle$ |
| $^3P_1$ | 20,280 – 21,000 | 20,584 | 20,724 | $+1.000\lvert ^3P_1\rangle$ |
| $^3P_2$ | 21,100 – 22,150 | 21,727 | 21,945 | $+0.955\lvert ^3P_2\rangle + 0.294\lvert ^1D_2\rangle$ |
| $^1S_0$ | – | – | 47,360 | $-0.996\lvert ^1S_0\rangle + 0.094\lvert ^3P_0\rangle$ |

The set of experimental barycenter energies was used to optimize wavefunctions for the $Pr^{3+}$-doped $Lu_2S_3$. The procedure described in detail by Hehlen *et al* [40] and the RELIC 1.0 software package documentation [41] was followed. Fitting the barycenter energies calculated from the optimized Hamiltonian matrix to a set of experimentally determined $Pr^{3+}$ barycenters, electrostatic interaction parameters $F_{(2)} = (296.0 \pm 2.4)$ $cm^{-1}$, $F_{(4)} = (46.0 \pm 1.2)$ $cm^{-1}$, $F_{(6)} = (4.66 \pm 0.17)$ $cm^{-1}$, and spin-orbit interaction parameter $\zeta = (739 \pm 13)$ $cm^{-1}$ can be obtained. Using these material-specific parameters optimized for $Pr:Lu_2S_3$, we calculated the position of the barycenters for all $Pr^{3+}$ levels and a complete set of intermediate coupling wavefunctions to see the impact of spin-orbit interaction (see Tab. 1).

It should be noted, the obtained interaction parameters $F_{(2,\ 4,\ 6)}$ and $\zeta$ for $Pr:Lu_2S_3$ are not significantly different from the average values of a large number of rare earth doped compounds (for $Pr^{3+}$:$F_{(2)} = 306$ $cm^{-1}$, $F_{(4)} = 46.8$ $cm^{-1}$, $F_{(6)} = 4.57$ $cm^{-1}$, $\zeta = 746$ $cm^{-1}$ [40,41]) and one can expect that tensor operators $\mathbf{U}^{(k)}$ and $\mathbf{L} + g\mathbf{S}$ for $Pr:Lu_2S_3$ will exhibit minimal discrepancy from the data available in the literature (e.g. [42]). This is basically true, but we have found that for some transitions (e.g. $^3F_4 \leftrightarrow {}^3H_4$, $^1G_4 \leftrightarrow {}^3H_4$, $^1D_2 \leftrightarrow {}^3H_4$, $^3F_4 \leftrightarrow {}^3H_5$, $^3F_4 \leftrightarrow {}^3H_6$, $^1G_4 \leftrightarrow {}^3H_6$, $^1G_4 \leftrightarrow {}^3H_5$, and $^1D_2 \leftrightarrow {}^3F_4$), notable discrepancies of the order of tens percent can be observed. For that reason, we also present (for all $Pr^{3+}$ transitions) the squared reduced-matrix elements for the tensor operators $\mathbf{U}^{(k)}$ and $\mathbf{L} + g\mathbf{S}$, calculated with the help of $F_{(2,\ 4,\ 6)}$ and $\zeta$ interaction parameters optimized for $Pr:Lu_2S_3$ (see Table. 2). These results were used to help identify transitions associated to some spectral lines in the measured $Pr:Lu_2S_3$ luminescence spectra. Furthermore, the results may be employed in a Judd-Ofelt analysis in the future once the $Pr:Lu_2S_3$ absorption cross-section data is available.

**Table 2**
Calculated set of optimized squared reduced-matrix elements for tensor operators $\mathbf{U}^{(k)}$ and $\mathbf{L} + g\mathbf{S}$.

| Transition | $\lambda$ [$\mu$m] | $\lvert\langle U^{(2)}\rangle\rvert^2$ | $\lvert\langle U^{(4)}\rangle\rvert^2$ | $\lvert\langle U^{(6)}\rangle\rvert^2$ | $\lvert\langle L + gS\rangle\rvert^2$ |
|---|---|---|---|---|---|
| $^3H_5 \leftrightarrow {}^3H_4$ | 4.81 | 0.1096 | 0.2007 | 0.6123 | 10.486 |
| $^3H_6 \leftrightarrow {}^3H_5$ | 4.62 | 0.1080 | 0.2327 | 0.6418 | 10.798 |
| $^3H_6 \leftrightarrow {}^3H_4$ | 2.36 | 0.0001 | 0.0338 | 0.1390 | 0 |
| $^3F_2 \leftrightarrow {}^3H_6$ | 22.5 | 0 | 0.0164 | 0.3044 | 0 |
| $^3F_2 \leftrightarrow {}^3H_5$ | 3.83 | 0 | 0.2979 | 0.6594 | 0 |
| $^3F_2 \leftrightarrow {}^3H_4$ | 2.13 | 0.5090 | 0.4032 | 0.1173 | 0 |
| $^3F_3 \leftrightarrow {}^3F_2$ | 7.22 | 0.0210 | 0.0510 | 0 | 6.513 |
| $^3F_3 \leftrightarrow {}^3H_6$ | 5.46 | 0 | 0.3181 | 0.8457 | 0 |
| $^3F_3 \leftrightarrow {}^3H_5$ | 2.50 | 0.6286 | 0.3468 | 0 | 0 |
| $^3F_3 \leftrightarrow {}^3H_4$ | 1.65 | 0.0653 | 0.3461 | 0.6980 | 0.007 |
| $^3F_4 \leftrightarrow {}^3F_3$ | 21.9 | 0.0255 | 0.0737 | 0.0053 | 4.367 |
| $^3F_4 \leftrightarrow {}^3F_2$ | 5.43 | 0.0012 | 0.0012 | 0.0907 | 0 |
| $^3F_4 \leftrightarrow {}^3H_6$ | 4.37 | 0.5688 | 0.6130 | 0.4616 | 0 |
| $^3F_4 \leftrightarrow {}^3H_5$ | 2.25 | 0.0295 | 0.3129 | 0.4399 | 0.167 |
| $^3F_4 \leftrightarrow {}^3H_4$ | 1.53 | 0.0193 | 0.0505 | 0.4917 | 0.171 |
| $^1G_4 \leftrightarrow {}^3F_4$ | 3.42 | 0.0789 | 0.1445 | 0.3490 | 2.689 |
| $^1G_4 \leftrightarrow {}^3F_3$ | 2.96 | 0.0039 | 0.0055 | 0.0524 | 2.377 |
| $^1G_4 \leftrightarrow {}^3F_2$ | 2.10 | 0.0000 | 0.0161 | 0.0063 | 0 |
| $^1G_4 \leftrightarrow {}^3H_6$ | 1.92 | 0.2568 | 0.2568 | 0.2432 | 0 |
| $^1G_4 \leftrightarrow {}^3H_5$ | 1.36 | 0.0380 | 0.0985 | 0.4205 | 0.147 |
| $^1G_4 \leftrightarrow {}^3H_4$ | 1.06 | 0.0014 | 0.0065 | 0.0230 | 0.060 |
| $^1D_2 \leftrightarrow {}^1G_4$ | 1.43 | 0.2956 | 0.0518 | 0.0755 | 0 |
| $^1D_2 \leftrightarrow {}^3F_4$ | 1.01 | 0.6051 | 0.0000 | 0.0200 | 0 |
| $^1D_2 \leftrightarrow {}^3F_3$ | 0.96 | 0.0327 | 0.0185 | 0 | 0.148 |
| $^1D_2 \leftrightarrow {}^3F_2$ | 0.85 | 0.0140 | 0.0872 | 0 | 0.078 |
| $^1D_2 \leftrightarrow {}^3H_6$ | 0.82 | 0 | 0.0707 | 0.0061 | 0 |
| $^1D_2 \leftrightarrow {}^3H_5$ | 0.70 | 0 | 0.0023 | 0.0003 | 0 |
| $^1D_2 \leftrightarrow {}^3H_4$ | 0.61 | 0.0031 | 0.0172 | 0.0541 | 0 |
| $^3P_0 \leftrightarrow {}^1D_2$ | 2.73 | 0.0154 | 0 | 0 | 0 |
| $^3P_0 \leftrightarrow {}^1G_4$ | 0.94 | 0 | 0.0558 | 0 | 0 |
| $^3P_0 \leftrightarrow {}^3F_4$ | 0.74 | 0 | 0.1067 | 0 | 0 |
| $^3P_0 \leftrightarrow {}^3F_3$ | 0.71 | 0 | 0 | 0 | 0 |
| $^3P_0 \leftrightarrow {}^3F_2$ | 0.65 | 0.2951 | 0 | 0 | 0 |
| $^3P_0 \leftrightarrow {}^3H_6$ | 0.63 | 0 | 0 | 0.0727 | 0 |
| $^3P_0 \leftrightarrow {}^3H_5$ | 0.55 | 0 | 0 | 0 | 0 |
| $^3P_0 \leftrightarrow {}^3H_4$ | 0.50 | 0 | 0.1729 | 0 | 0 |
| $^1I_6 \leftrightarrow {}^3P_0$ | 869 | 0 | 0 | 0.0031 | 0 |
| $^1I_6 \leftrightarrow {}^1D_2$ | 2.72 | 0 | 0.1554 | 1.7042 | 0 |
| $^1I_6 \leftrightarrow {}^1G_4$ | 0.94 | 0.2426 | 1.3860 | 0.6684 | 0 |

### *3.4. Luminescence spectra*

In order to characterize the $Pr:Lu_2S_3$ luminescence spectra in detail, and to assign the observed spectral lines to the corresponding transitions, several excitation wavelengths were employed to selectively excite the main energy levels (namely $^3P_2$, $^1D_2$, $^1G_4$, and $^3F_4$) of the $Pr:Lu_2S_3$ material. All measurements were carried out at room temperature. In total, the obtained luminescence spectra of the $Pr:Lu_2S_3$ crystal covered the wavelength range from 460 nm to 5.4 $\mu$m.

#### *3.4.1. Excitation of $^3P_2$ energy level*

The $^3P_2$ level was excited by radiation with a wavelength of 456 nm. During this excitation, no emission at a wavelength below 497 nm was observed, as shown in the luminescence spectra in Fig. 6. Taking into account the proximity of the $^3P_2$, $^3P_1$ and $^3P_0$ levels, the majority of the $Pr^{3+}$ ions are expected to relax rapidly to the $^3P_0$ or $^1I_6$ state. Furthermore, considering the size of the squared reduced-matrix elements for the tensor operators $\mathbf{U}^{(k)}$, it can be assumed that most of the observed emission lines correspond to transitions originating from the $^3P_0$ level. However, to explain the origin of some emission lines, it was necessary to acknowledge that the starting level is also $^1I_6$ (transitions $^1I_6 \rightarrow {}^3H_5$, $^1I_6 \rightarrow {}^3F_3$, $^1I_6 \rightarrow {}^1G_4$), since the corresponding transitions starting from the $^3P_0$ level are either forbidden or very unlikely, see Tab. 2 (from this point of view it is noteworthy that although the values of $\lvert\langle \mathbf{U}^{(2,4,6)}\rangle\rvert^2$ for $^1I_6 \rightarrow {}^1G_4$ transition belongs to the largest, the corresponding emission is rather less intense).

It should be noted that upon excitation of the $^3P_2$ level, spectral lines associated with $^1D_2$ level transitions are also observable in the emission spectrum, see Fig. 6. This is evident by comparison of that spectrum with the spectrum obtained upon direct excitation of the $^1D_2$ level, as described in 3.4.2 (Fig. 7).

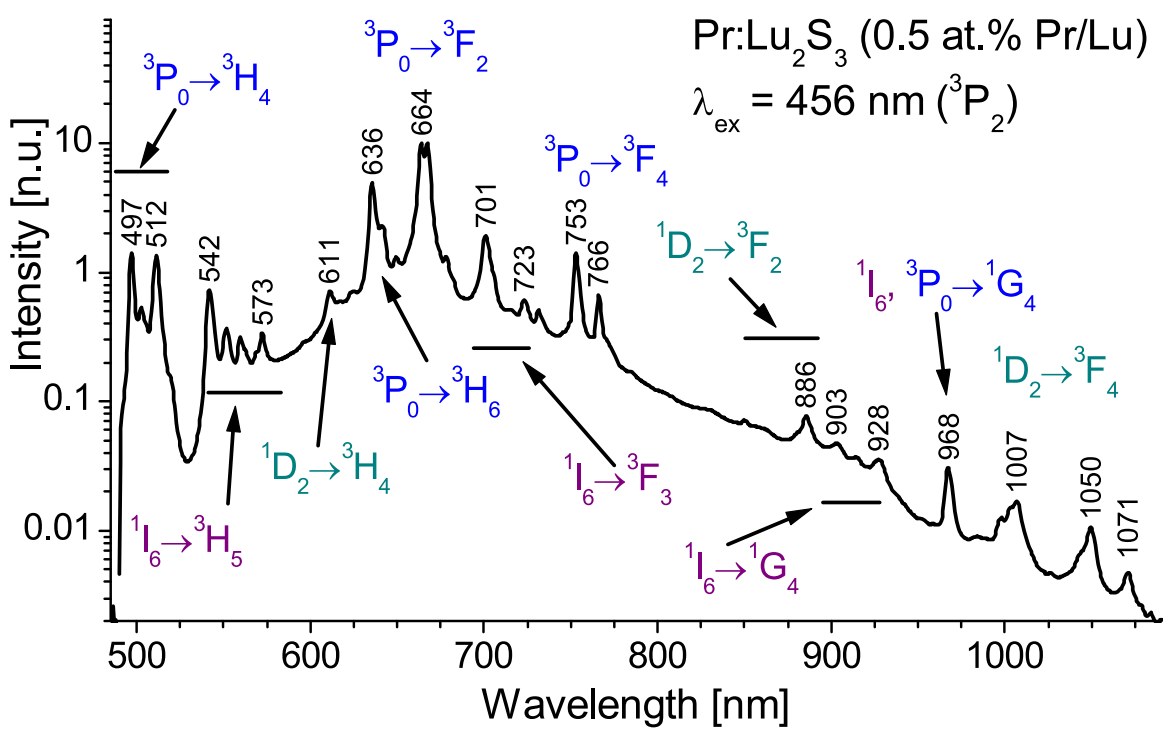

**Fig. 6.** Luminescence spectra of Pr:$Lu_2S_3$ corresponding to $^3P_2$ excitation at wavelength 456 nm measured by BlueWave spectrometer in range 480 – 1080 nm. Transitions with a different starting energy level are distinguished by a different color.

It is also worth mentioning, the luminescence spectra (see Fig. 6 and Fig. 7a) are distorted by the luminescence signal originating from the color centers in the crystal which are characterized by the broad emission, typically in the visible spectral range, resulting in a pedestal-like background in the measured spectra. The precise nature of the color center has not yet been investigated. However, sulfur vacancies and the substitution of sulfur by oxygen are considered the most likely origins.

### 3.4.2. *Excitation of $^1D_2$ energy level*

The luminescence spectra obtained upon excitation of the $^1D_2$ level by the 596 nm wavelength radiation are depicted in Fig. 7, along with the respective transitions. Besides the expected spectral lines originating from the $^1D_2$ level, followed by the transitions from the subsequent lower energy levels ($^1G_4$, $^3F_4$, and $^3F_3$), there are some lines in the spectrum (namely 664, 753, and 766 nm) which can not be assigned to any of these transitions (based on the calculated squared reduced-matrix elements for the tensor operators $\mathbf{U}^{(k)}$, see Tab. 2). Nevertheless, these lines can be well explained by transitions from the $^3P_0$ level. Therefore, the hypothesis is that there exists some energy transfer processes, such as the absorption of excited states (e.g. $^3H_6 \rightarrow {}^3P_1$ or $^3F_3 \rightarrow {}^3P_2$), which populate the $^3P_0$ level.

### 3.4.3. *Excitation of $^1G_4$ energy level*

In the case of the $^1G_4$ level excitation, radiation at 1022 nm wavelength was used. The corresponding luminescence spectra recorded in the range of 1200 – 1700 nm, 1750 – 2040 nm, and 2000 – 3600 nm are shown in Fig. 8a, b, and Fig. 8c, respectively.

Note, the spectrum assigned to the $^1G_4 \rightarrow {}^3H_5$ transition exactly matches the corresponding part of the spectrum observed upon excitation of the $^1D_2$ level (see Fig. 7b). If we assume that the sensitivity of the dedicated spectrometer does not vary significantly within the specified measuring spectral range, it can be reasonably deduced (from Fig. 7b) that the majority (approx. 75 %) of $Pr^{3+}$ ions excited to the level $^1G_4$ will subsequently leave this level directly via the channel $^1G_4 \rightarrow {}^3H_5$. This deduction is based on the ratio of the integrated parts of the luminescence spectrum shown in Fig. 7b, which are associated with the transitions $^1G_4 \rightarrow {}^3H_5$ and $^1D_2 \rightarrow {}^1G_4$. From the data presented in Fig. 8a, we know the exact spectrum corresponding to the pure $^1G_4 \rightarrow {}^3H_5$ transition. The integral over this spectrum, covering the 1270 – 1410 nm range, is 25.5 au · nm, and this value is proportional to the number of $^1G_4 \rightarrow {}^3H_5$ transitions. We expect that the luminescence within the 1410 – 1610 nm spectral range (see Fig. 7b) corresponds to the $^1D_2 \rightarrow {}^1G_4$ transition, with the respective integral of 34 au · nm. This value is proportional to the number of $^1D_2 \rightarrow {}^1G_4$ transitions. The ratio of these integrals is 0.75, meaning that 75 % of the photons reaching the $^1G_4$ level via the $^1D_2 \rightarrow {}^1G_4$ transition leave this level through the $^1G_4 \rightarrow {}^3H_5$ channel.

Furthermore, luminescence around a wavelengths of 1.9 $\mu$m, 3 $\mu$m, and 3.4 $\mu$m which can be assigned to the $^1G_4 \rightarrow {}^3H_6$, $^1G_4 \rightarrow {}^3F_3$ and $^1G_4 \rightarrow {}^3F_4$ transitions, respectively, was observed. It should be remarked that the recorded spectrum corresponding to $^1G_4 \rightarrow {}^3F_4$ transition was constrained by the range of the PbS detector used. Therefore, it is possible that this emission may extend further into the infrared region.

### 3.4.4. *Excitation of $^3F_4$ energy level*

The Pr:$Lu_2S_3$ emission recorded upon excitation of the $^3F_4$ and $^3F_3$ levels is also of great interest, because it covers wide spectral bands in the infrared region with the potential to generate radiation of up to ~7 $\mu$m wavelength. A large number of sublevels corresponding to $^3F_4$ and $^3F_3$ levels provide a series of relatively broad absorption bands from 1.4 to 1.7 $\mu$m (see Fig. 5b), which can be easily excited by many commercially available pumping sources, like laser diodes, Er-doped fiber or Raman fiber lasers.

The luminescence spectra obtained upon excitation of the $^3F_4$ level by 1453 nm wavelength radiation are displayed in Fig. 9, for the spectral range of the 1500 nm – 1700 nm ($^3F_{4,3} \rightarrow {}^3H_4$ transition).

Subsequently, radiation with a wavelength of 1556 nm was used to excite the $^3F_{4,3}$ level, which provided a more intense luminescence signal. With the help of a monochromator along with the sensitive detectors, the emission spectra in the range of 1.9 – 2.8 $\mu$m (Fig. 10a) and 4 – 5.4 $\mu$m (Fig. 10b) were obtained. It should be remarked, that the part of the spectrum above 5 $\mu$m (Fig. 10b) is affected by the decreasing sensitivity of the PbSe detector.

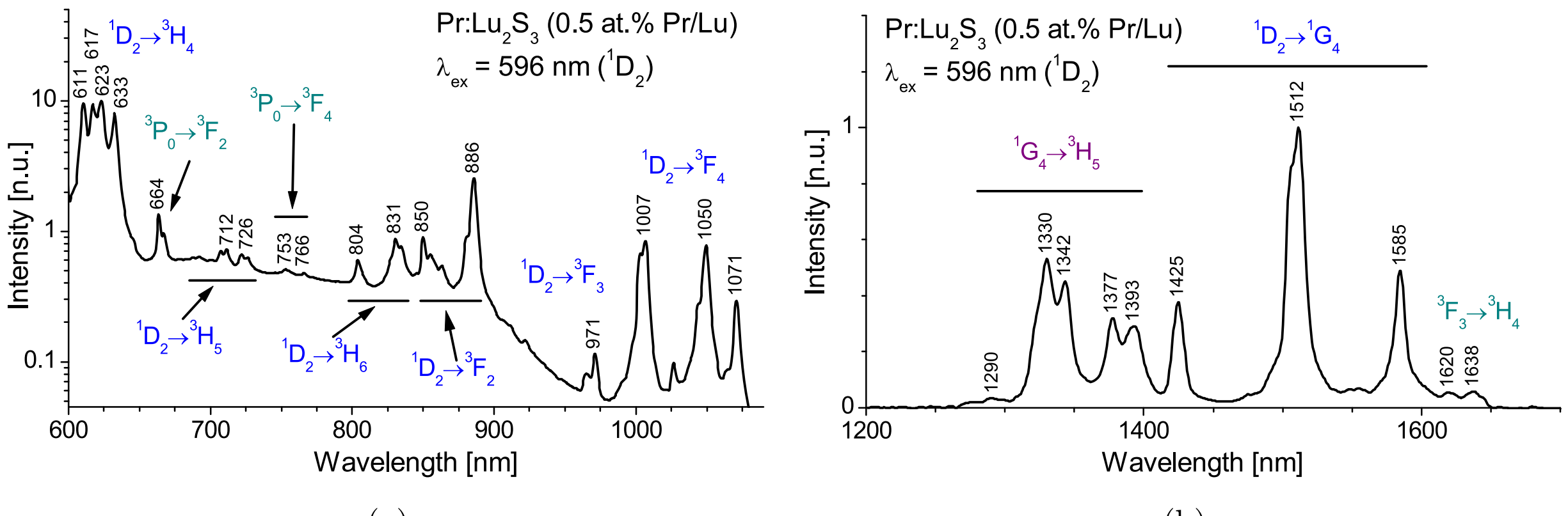

(a) (b)

**Fig. 7.** Luminescence spectra of Pr:$Lu_2S_3$ corresponding to $^1D_2$ excitation at wavelength 596 nm measured by BlueWave spectrometer in range 600 – 1080 nm (a) and by DwarfStar spectrometer in range 1200 – 1700 nm (b). Transitions with a different starting energy level are distinguished by a different color.

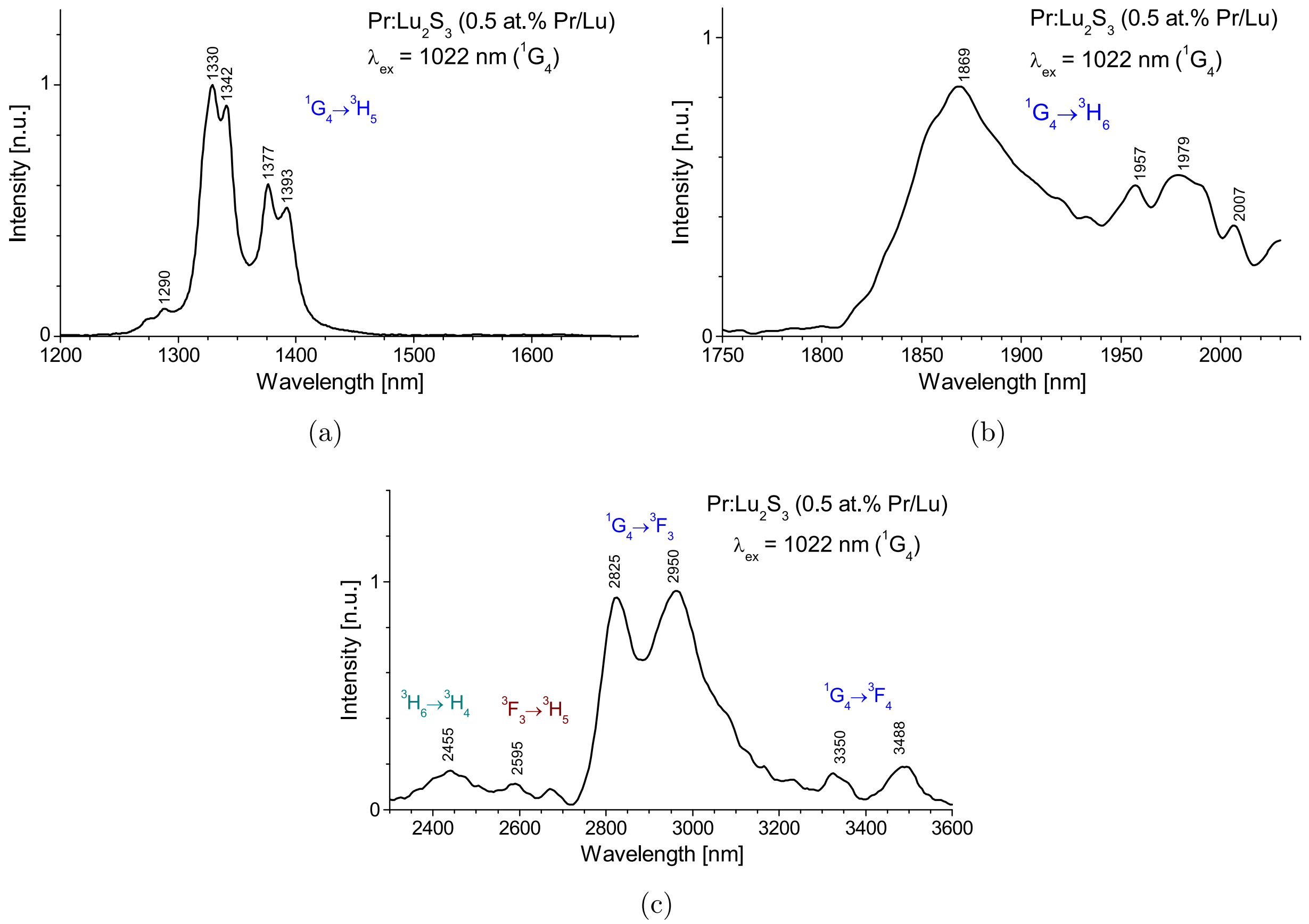

**Fig. 8.** Luminescence spectra of $Pr:Lu_2S_3$ corresponding to $^1G_4$ excitation at wavelength 1022 nm measured behind the 1.2 $\mu$m long-pass filter (FELH1200, Thorlabs) by DwarfStar spectrometer in range 1.2 – 1.7 $\mu$m (a), by Oriel monochromator (slit width 1.3 mm) in combination with PbS detector and long-pass filter (FEL 1500, Thorlabs) in range 1.75 – 2.04 $\mu$m (b), and by Oriel monochromator (slit width 2 mm) along with PbS detector and 2.05 $\mu$m long-pass filter (LP-2050, Spectrogon) in range 2.3 – 3.6 $\mu$m (c). Transitions with a different starting energy level are distinguished by a different color.

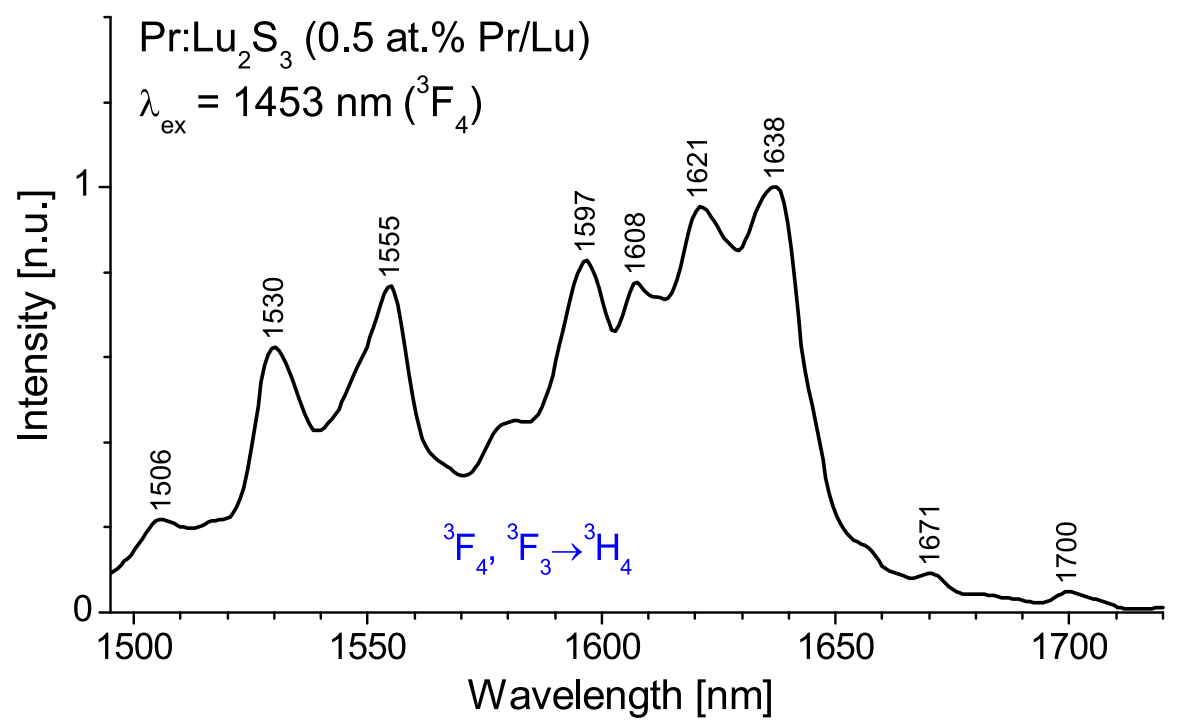

**Fig. 9.** Luminescence spectra of $Pr:Lu_2S_3$ corresponding to $^3F_4$ excitation at wavelength 1453 nm and subsequent transition $^3F_{4,3} \rightarrow {}^3H_4$ measured by DwarfStar spectrometer in range 1500 – 1700 nm behind the 1.5 $\mu$m long-pass filter (FELH1500, Thorlabs).

Moreover, it can be reasonably assumed that upon excitation of the $^3F_{4,3}$ level, luminescence transitions even at $^3F_3 \rightarrow {}^3H_6$ ($\sim$ 5.5 $\mu$m) and $^3F_3 \rightarrow {}^3F_2$ ($\sim$ 7 $\mu$m) can be reached from the $Pr:Lu_2S_3$ material. This assumption is based on the following deduction: the distance between the $^3F_2$ and $^3F_3$ level barycenters (described in subsection 3.3) is only $\sim$1400 cm$^{-1}$ (see Tab. 1), so the gap between the lowest $^3F_3$ sublevel and the highest $^3F_2$ sublevel will be even smaller, thus only 4 phonons may be enough to overcome it (given the estimated phonon energy of 312 cm$^{-1}$ for $Lu_2S_3$). For that reason, one could expect significant non-radiative depopulation of the $^3F_3$ level. However, the observed luminescence from this level is proof that $^3F_3$ is only weakly quenched by multiphonon relaxation in the case of $Pr:Lu_2S_3$. Therefore, it is legitimate to expect that with appropriate detector it would be possible to observe emission even at wavelengths of up to 7 $\mu$m.

## 4. Conclusion

Spectroscopic properties of a new low-phonon material, $Pr:Lu_2S_3$, prepared by a modified micro-pulling-down method, were presented. The $Pr:Lu_2S_3$ was investigated for its potential application as a low-phonon solid state laser material for mid-IR laser sources; excitation and luminescence spectra up to 5.4 $\mu$m were described, for the first time to our best knowledge. To characterize the luminescence spectra of the $Pr:Lu_2S_3$ in detail, several excitation wavelengths were used to excite the main energy levels of the $Pr^{3+}$-ion; doing that along with the help of the calculated squared reduced-matrix elements, 26 transitions associated to the measured emission spectra were identified in the energy level structure of the $Pr:Lu_2S_3$.

In addition, calculated barycenters, intermediate coupling wave-functions, and optimized squared reduced-matrix elements for tensor operators $\mathbf{U}^{(k)}$ and $\mathbf{L} + g\mathbf{S}$ are presented in this contribution for the $Pr:Lu_2S_3$ material. Our next goal is to use these values for the Judd-Ofelt analysis in the future, to obtain even more information about this crystal, providing, the quality of the grown sample will be enough to determine the $Pr:Lu_2S_3$ absorption cross-sections.

It should be noted that the poor quality of the grown sample, which is limited by the current material preparation technology, prevents the acquisition of relevant data necessary to determine the fluorescence lifetime. Defects and color centers in the crystal may act as trapping centers or contribute to quenching mechanisms, leading to potentially misleading results. Once improvements are made in the growth technology (e.g., by selecting a more appropriate crucible, as described in 2.1) and higher-quality samples become available, we plan to build on the existing knowledge and fully spectroscopically characterize the crystal.

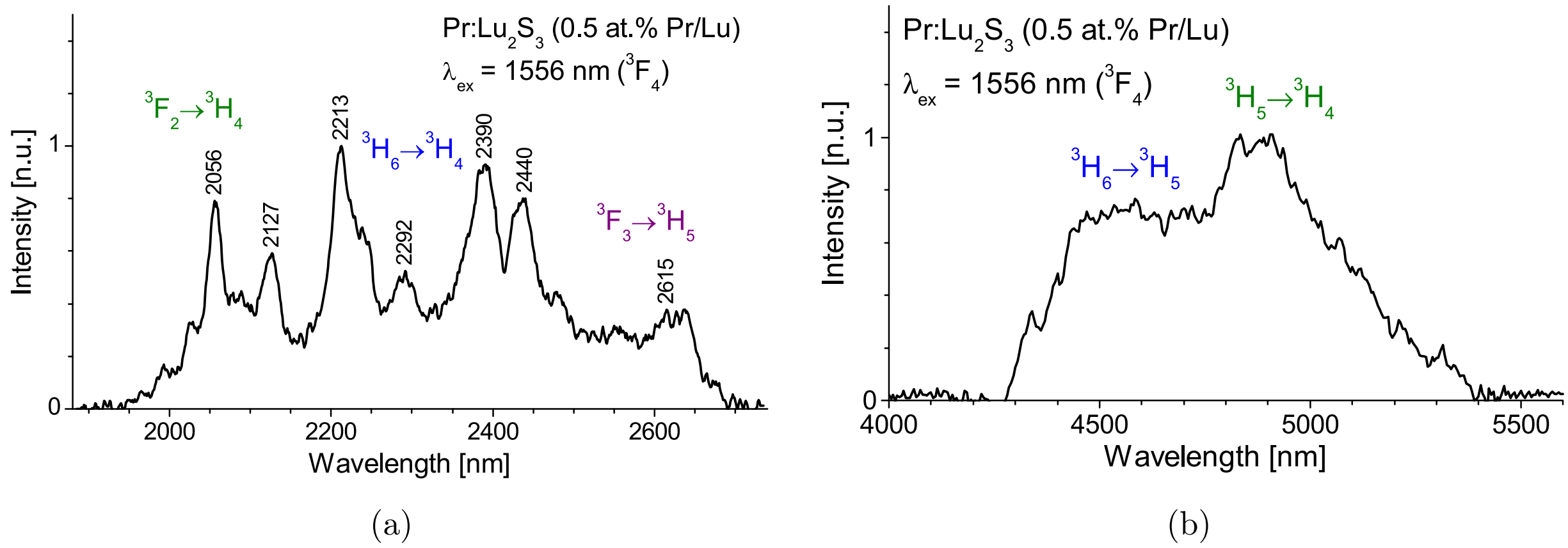

**Fig. 10.** Luminescence spectra of Pr:$Lu_2S_3$ for the spectral range of 1.9 − 2.8 μm (monochromator Oriel, slit width 1 mm, PbS detector and long-pass filter LP-1850, Spectrogon) (a) and 4 − 5.5 μm (monochromator Oriel, slit width 8 mm, PbSe detector and long-pass filter LP-3000, Spectrogon) (b). Transitions with a different starting energy level are distinguished by a different color.

It should be also remarked that although we cannot directly compare (as explained earlier in the text) $Lu_2S_3$ properties such as lifetimes, effective cross-sections, and/or multi-phonon relaxation rates with those of other well-established low-phonon laser hosts, we believe the host matrix itself has no direct competitor among the low-phonon materials. We are not aware of any matrices that are simultaneously comparable to $Lu_2S_3$ in all key parameters, such as being low-phonon, transparent in the deep infrared region, non-hygroscopic, and tolerant to high doping concentrations of lanthanide ions. In this regard, the $Lu_2S_3$ is unique. The only comparable material is the $KLuS_2$ crystal, but it has not yet been successfully grown in bulk form.

So, based on the presented results, it can be concluded that Pr:$Lu_2S_3$ material has the potential to be an interesting candidate for mid-infrared laser sources. In addition, the $Lu_2S_3$ matrix is definitely attractive for further lanthanide doping.

## CRediT authorship contribution statement

**Martin Fibrich:** Writing – original draft, Visualization, Methodology, Investigation, Formal analysis, Data curation, Conceptualization. **Lubomír Havlák:** Writing – review & editing, Validation, Methodology, Investigation, Conceptualization. **Jan Šulc:** Writing – review & editing, Visualization, Validation, Methodology, Investigation, Formal analysis, Data curation, Conceptualization. **Robert Král:** Writing – review & editing, Validation, Methodology, Investigation, Conceptualization. **Vítězslav Jarý:** Writing – review & editing, Validation, Methodology, Conceptualization. **David Vyhlídal:** Writing – review & editing, Visualization, Software. **Vojtěch Vaněček:** Writing – review & editing, Methodology, Investigation, Conceptualization. **Martin Nikl:** Writing – review & editing, Supervision, Resources, Funding acquisition. **Helena Jelínková:** Writing – review & editing, Supervision, Resources, Project administration, Funding acquisition, Conceptualization.

## Funding

The work is supported by Operational Programme Johannes Amos Comenius financed by European Structural and Investment Funds and the Czech Ministry of Education, Youth and Sports (Project LASCIMAT – CZ.02.01.01/00/23_020/0008525).

## Declaration of Competing Interest

The authors declare that they have no known competing financial interests or personal relationships that could have appeared to influence the work reported in this paper.